\documentclass[12pt,preprint]{aastex}
\usepackage{psfig}
\begin{document}
\title{Is the Pre-WMAP CMB Data Self-consistent?}
\author{Charles H. Lineweaver and Louise M. Griffiths\\
Department of Astrophysics and Optics, School of Physics\\
University of New South Wales, Sydney, NSW 2052, Australia}

\begin{abstract}
Although individual observational groups vigorously test their data
sets for systematic errors, the pre-WMAP CMB observational data set has
not yet been collectively tested.  Under the assumption that the
concordance model is the correct model, we have explored residuals of
the observational data with respect to this model to see if any
patterns emerge that can be identified with systematic errors.
We found no significant trends associated 
with frequency, frequency channels, calibration source, pointing uncertainty, instrument type, platform and 
altitude.
We did find some evidence at the $\sim 1$ to $\sim 2 \sigma$ level for trends associated with angular scale 
($\ell$ range) and absolute galactic latitude. The slope of the trend in galactic latitude is consistent
with low level galactic contamination. The residuals with respect to $\ell$ may indicate that the concordance model
used here needs slight modification. See Griffiths \& Lineweaver (2003) for more detail.
\end{abstract}


\section{Motivation and Method}

The ever-tightening network of constraints from CMB and non-CMB observations 
favors a concordant $\Lambda$ cold dark matter (CDM) model that is commonly 
accepted as the standard cosmological model. As long as the systematic errors
associated with CMB observations are small, the CMB power spectrum can play 
an increasingly large role in establishing and refining this model. Thus, it is 
crucial to check the CMB data for possible systematic errors in as many ways 
as possible.

Systematic errors and selection effects are notoriously difficult to
identify and quantify.  Calibration and/or beam uncertainties dominate
CMB measurements and there may be lower level systematic errors of which 
we are not aware.  Individual experimental groups have developed various 
ways to check their CMB observations for systematic effects
\cite[e.g.][]{kogut96,miller02},
including the use of multiple calibration sources, multiple frequency
channels and extensive beam calibrating observations. Internal
consistency is the primary concern of these checks.

Testing for consistency with other CMB observations is another
important way to identify possible systematic errors.  When the areas
of the sky observed overlap, this can be done by comparing CMB
temperature maps 
\cite[e.g.][]{ganga94,lineweaveretal95,xu01}. 
When similar angular scales are being observed one can compare power
spectra (e.g. Sievers et al. 2002).
and, for example, check for frequency-dependent discrepancies
(Odman 2003).

To calibrate a data set,  one needs a trusted external calibrator.
In this work we take the point of view that the $\Lambda$CDM concordance model,
that has emerged over the past 5 years due to its compatibility with dozens of observations with 
independent systematic errors, can be used as a calibrator.
These independent observations are the Hubble Key Project constraint on $h$ (Freedman et al 2001), the
$\Omega_m-\Omega_{\Lambda}$ constraints from observations of type Ia supernovae
(Riess et al. 1998, Perlmutter et al. 1999),
big bang nucleosynthesis constraints on the baryon content (Burles et al. 2001),
and large scale structure constraints on the amplitude $\sigma_8^2$ and shape of the
matter power spectrum from the 2-degree Field Galaxy Redshift Survey (Peacock et al. 2001).
Recent joint likelihood analyses of this data and the CMB (as summarized in Griffiths \& Lineweaver 2003) suggest the 
observationally concordant cosmology;
$\Omega_{\kappa}\simeq0$, $\Omega_{\Lambda}\simeq 0.7$ ($\Omega_{m} =
\Omega_b+\Omega_c\simeq 0.3$), $\Omega_b h^2 \simeq0.02$, $n_s \simeq
1$ and $h \simeq 0.68$ with $A_t$, $\tau$ and $\Omega_{\nu}$ taken to
be zero.

Our analysis is based on the assumption that the combined cosmological
observations used to determine the concordance model are giving us a
more accurate estimate of cosmological parameters, and therefore of
the true $C_{\ell}$ spectrum, than is given by the CMB data
alone. Under this assumption, the residuals of the individual observed
CMB band powers and the concordance $\Lambda$CDM model become tools to
identify a variety of systematic errors. 
We look for any linear trends that may identify systematic effects that are correlated
with the details of the observations.

A prerequisite for the
extraction of useful estimates for cosmological parameters from the
combined CMB data set is the mutual consistency of the observational
data points.
However, for the sake of a clean interpretational story, systematic errors in a data set are sometimes ignored or
explained away.  
David Wilkinson was fond of explicitly showing the systematic errors, or warts as he called 
them, in his data.
In a similar spirit we went looking for warts in the pre-WMAP data.

\begin{figure}[!ht]
\centerline{\psfig{figure=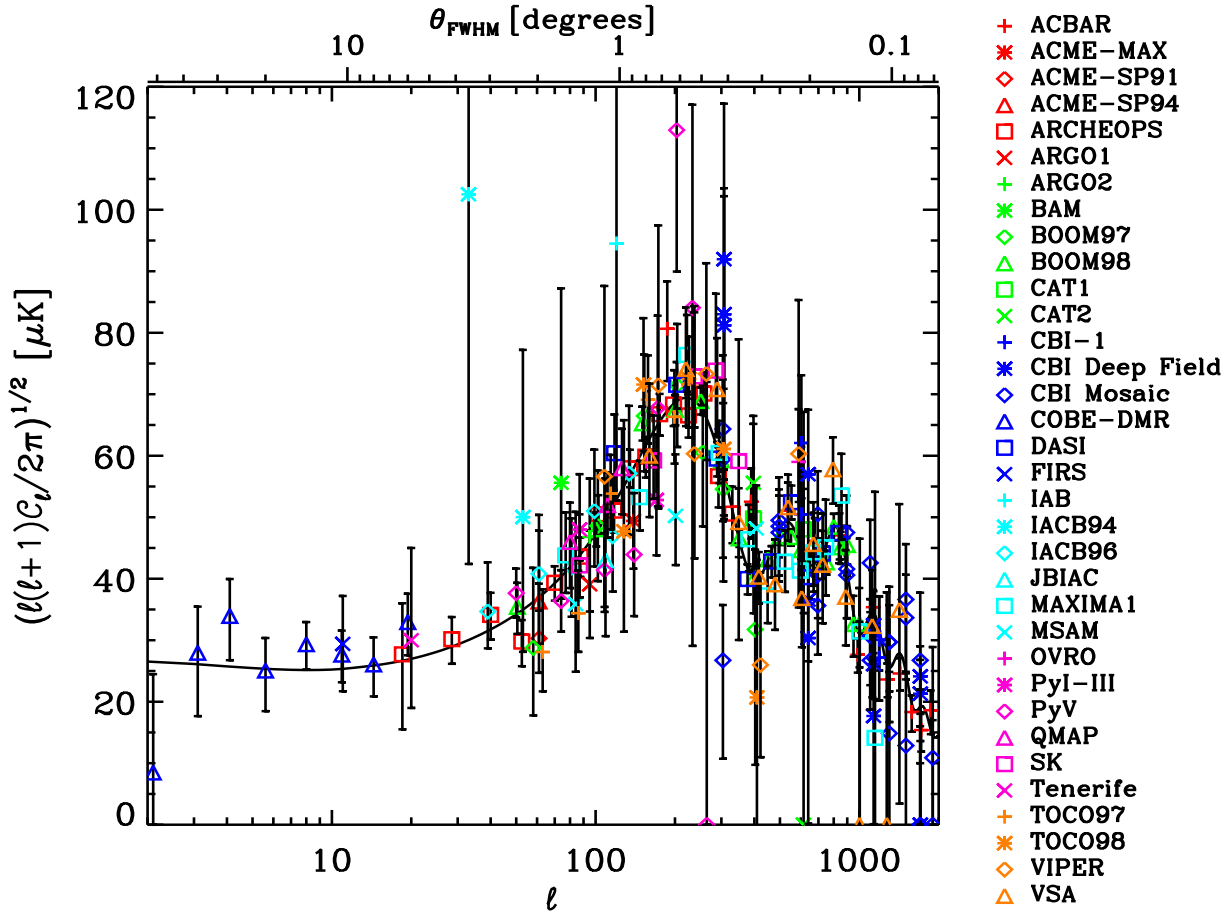,height=14cm,width=17.5cm}}
\vspace{-0.8cm}
\caption{ The concordance model and the pre-WMAP data. 
Even in this unbinned jumble the Sachs-Wolfe plateau ($2 \le \ell \le 40$) 
and the first two acoustic peaks can be seen.
The normalization of the model results from a $\chi^{2}$  minimization that includes 
marginalization over calibration uncertainties. 
We find $Q_{10} = 16.3 \pm 0.1 \;\mu$K, where
$10(10+1) C_{10} = \frac{24\pi}{5} Q_{10}^{2}$
(Lineweaver \& Barbosa 1998).
The very small error bar on the normalization of this model is due to the fact
that we have conditioned on values of other cosmological parameters that are usually marginalized over. 
The minimized $\chi^2$ for the concordance model is 174.2 for $\approx 175$
degrees of freedom. Thus we have a good fit.
Our $\chi^{2}$ analyses are performed on this raw, unbinned data but for clarity in the following figures, 
we show only the binned residuals.
See Griffiths \& Lineweaver (2003) for more details.
}
\label{clplotcol}
\end{figure}

\clearpage
\begin{figure}[!ht]
\centerline{\psfig{figure=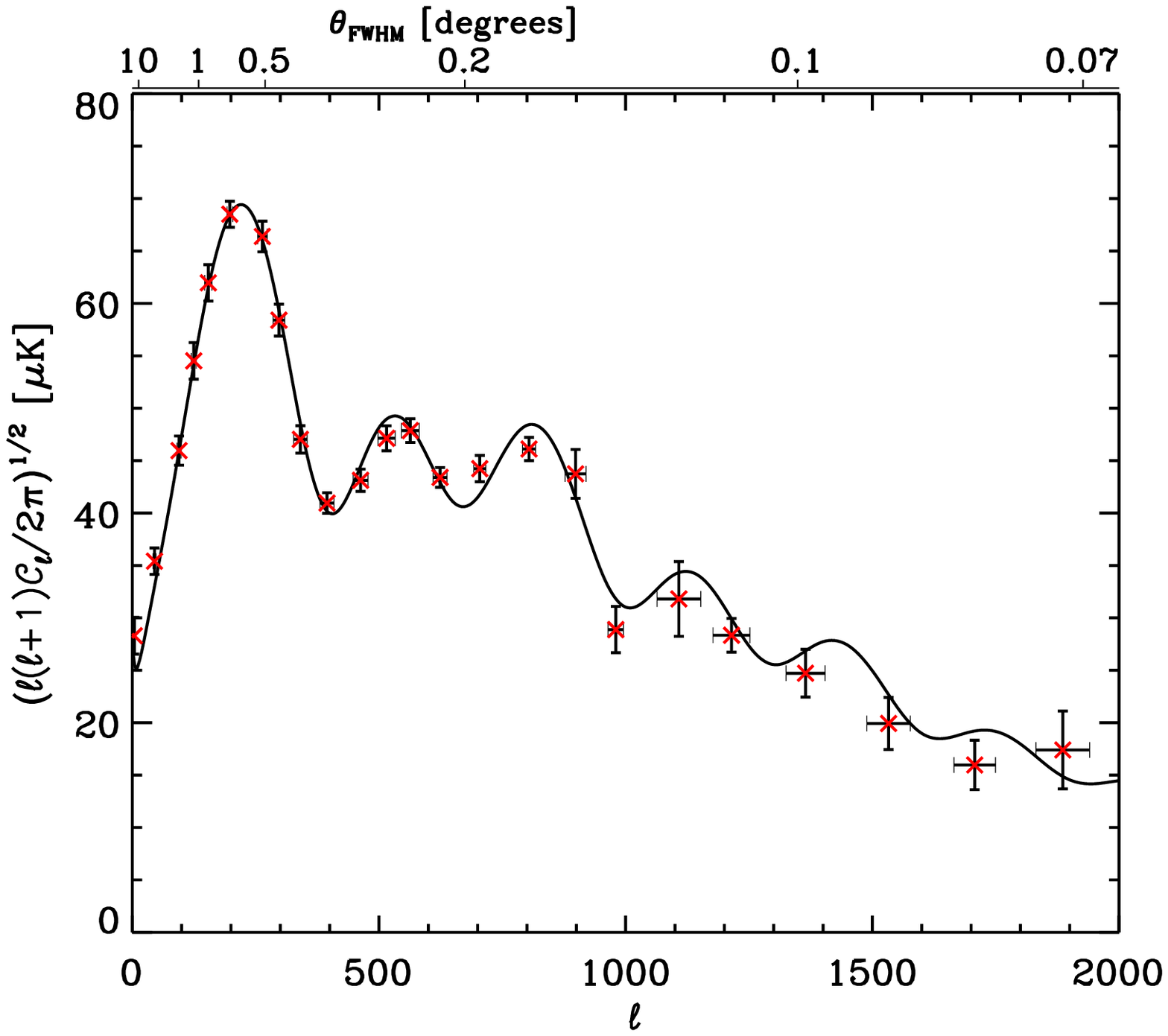,height=8.4cm,width=14cm}}
\vspace{-1.2cm}
\centerline{\psfig{figure=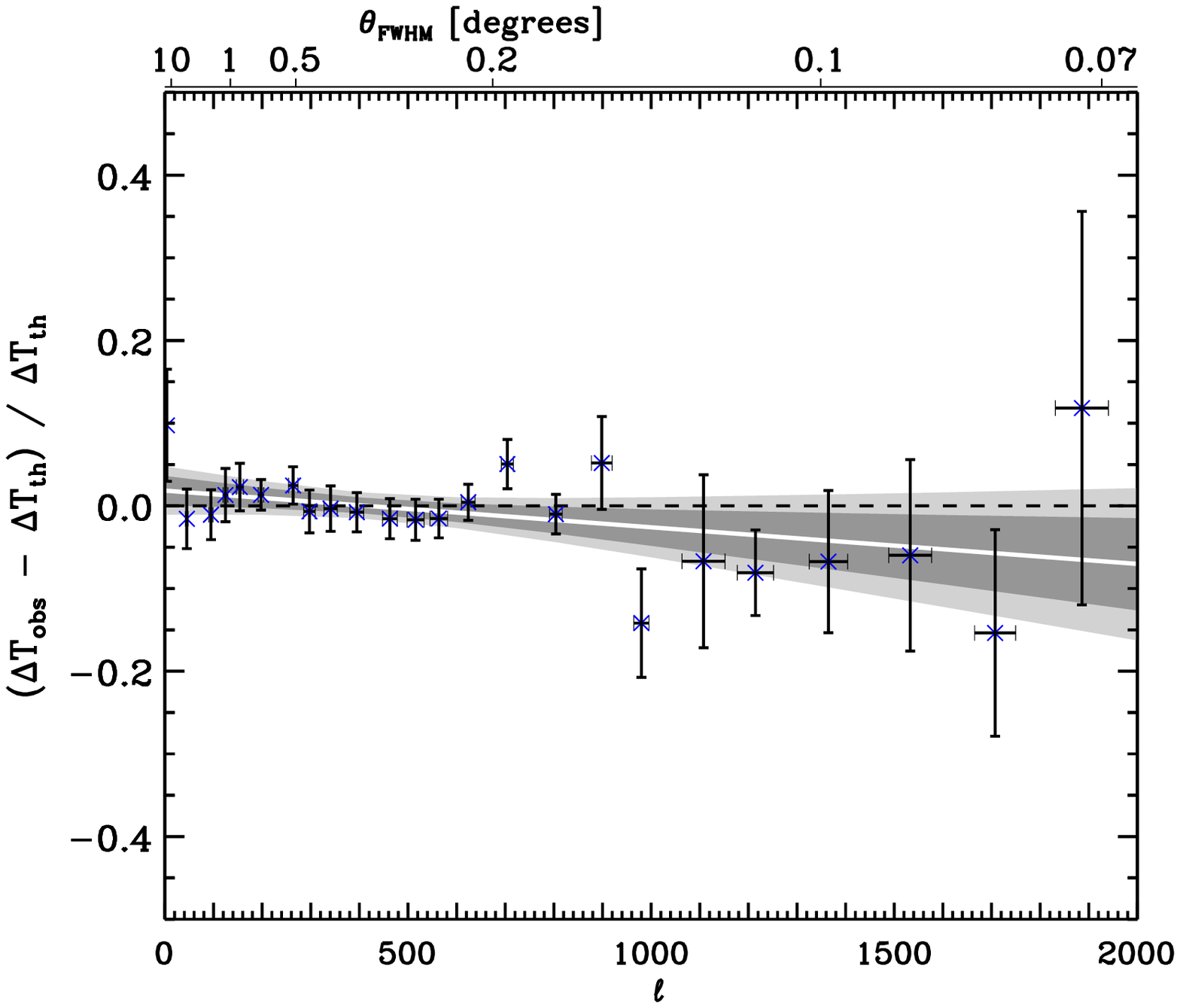,height=8.4cm,width=14cm}}
\vspace{-1cm}
\caption{TOP: Binned version on a linear scale of the previous plot showing the tightly constrained first 
two acoustic peaks, some evidence for a third and a general reduction of power as we move to smaller scales.
BOTTOM: residuals plotted against $\ell$. 
Notice that the residuals are not very randomly scattered around zero.
There is a run of 6 low points in the range ($900 < \ell < 1800$) and a pattern resembling a low amplitude 
sine wave in the range ($2 < \ell < 700$). These non-random patterns are telling us that
some new model may be required despite the fact that the $\chi^{2}$ shows the concordance model to be a good fit.
Shifting of the peak to the left or fattening the peak with a bit of $\tau$ would probably eliminate the sine wave.
The run of 6 points at high $\ell$ may be due to a systematic calibration error for some of the experiments in this
$\ell$ range and/or more suppression on small scales needs to be included in the concordance model.
The best-fit line to the data is shown in white and has a $\chi^2/DOF = 171/174 \approx 0.98$.
The best-fit line is surrounded by 68\% (dark grey) and 95\% (light grey) 
confidence intervals.
The concordance model (the dashed horizontal line) is $\sim 1.5 \sigma$
from the best fit.
}
\label{bin_clplotcol}
\end{figure}

\clearpage
\begin{figure}[!ht]
\centerline{\psfig{figure=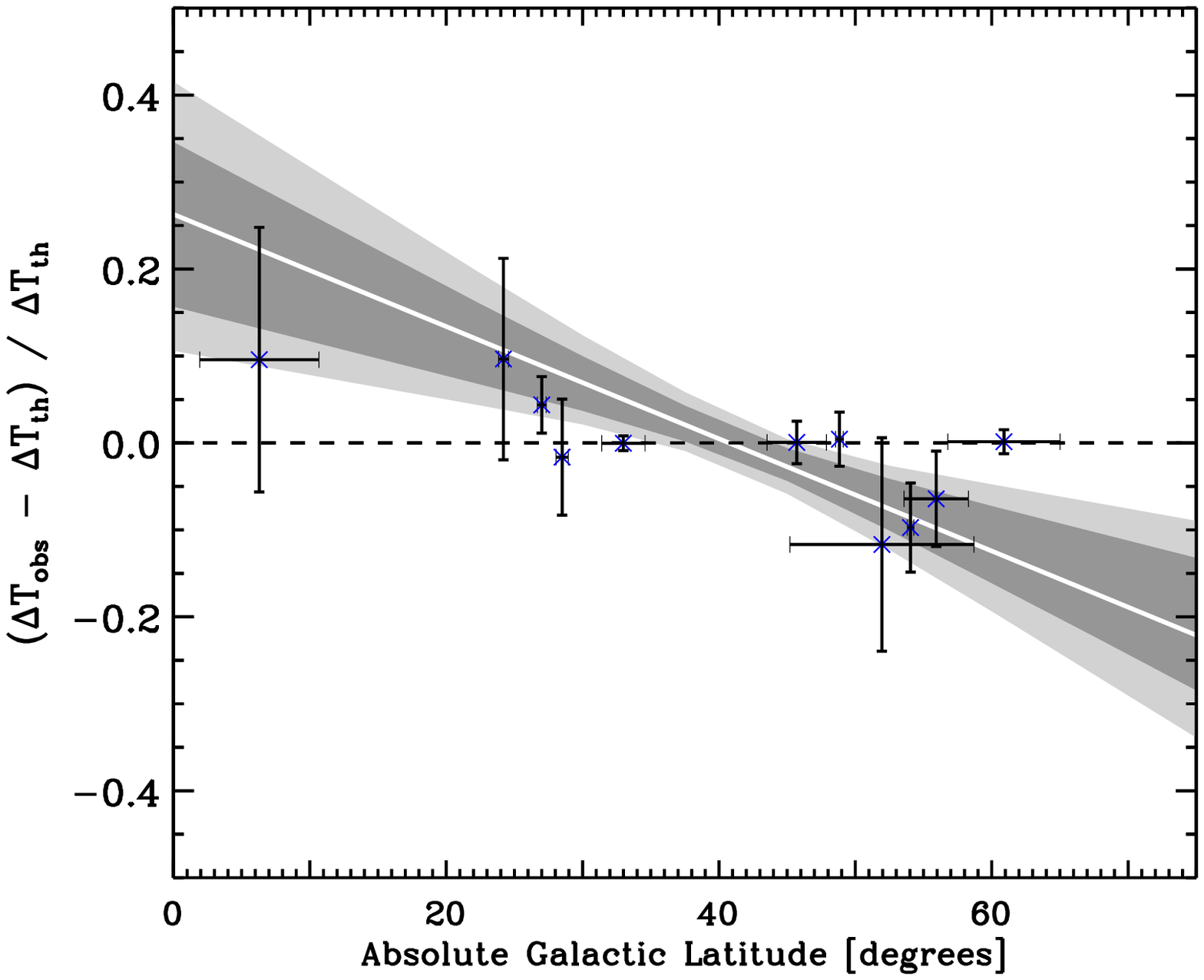,height=9.5cm,width=15cm}}
\vspace{-1.4cm}
\centerline{\psfig{figure=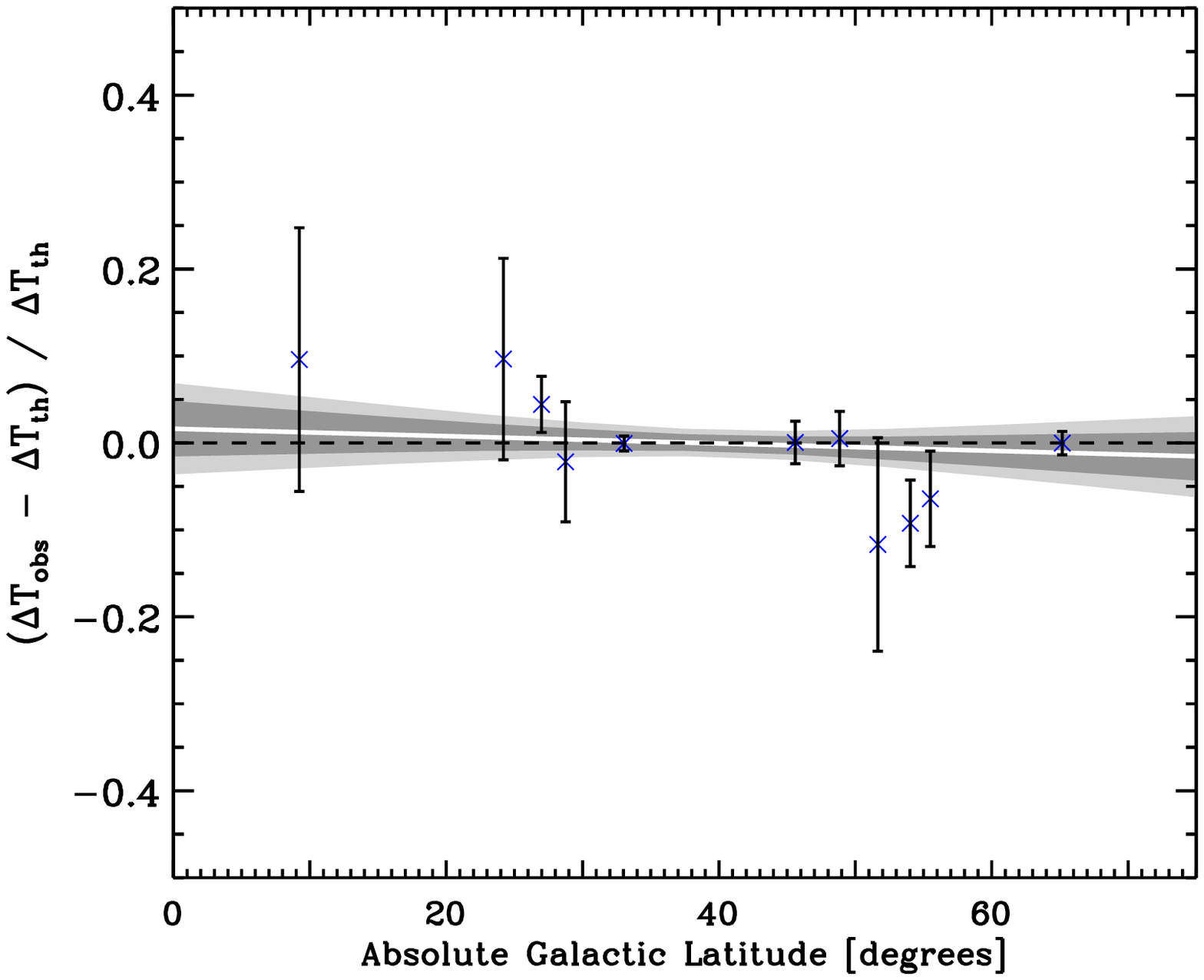,height=9.5cm,width=15cm}}
\vspace{-1.0cm}
\caption{
Observations taken at lower absolute galactic latitudes, $|b|$, will
be more prone to galactic contamination.
We check for this effect by
examining the residuals as a function of $|b|$.
TOP: the fitting routine 
includes the range in $|b|$ (horizontal error bars) as the uncertainty in $|b|$. The $\chi^2/DOF = 154/174 = 0.88$.
The probability of finding a line that better fits the data is 14\%.
The negative slope of the trend is consistent with low levels of galactic contamination.
BOTTOM: when we do not allow the data to shift horizontally, the best-fit
line is nearly flat. This is because the most distant outliers from the white line in the top
panel have large ranges of $|b|$.
The $\chi^2/DOF = 173.5/174 \approx 1$. We believe the most plausible result
is intermediate between these two cases since using ranges in $|b|$ as statistical
errors is problematic (top), but so too is treating the $|b|$ values as if they have no
errors (bottom).
}
\label{gallatresid}
\end{figure}

\clearpage
\begin{figure}[!ht]
\centerline{\psfig{figure=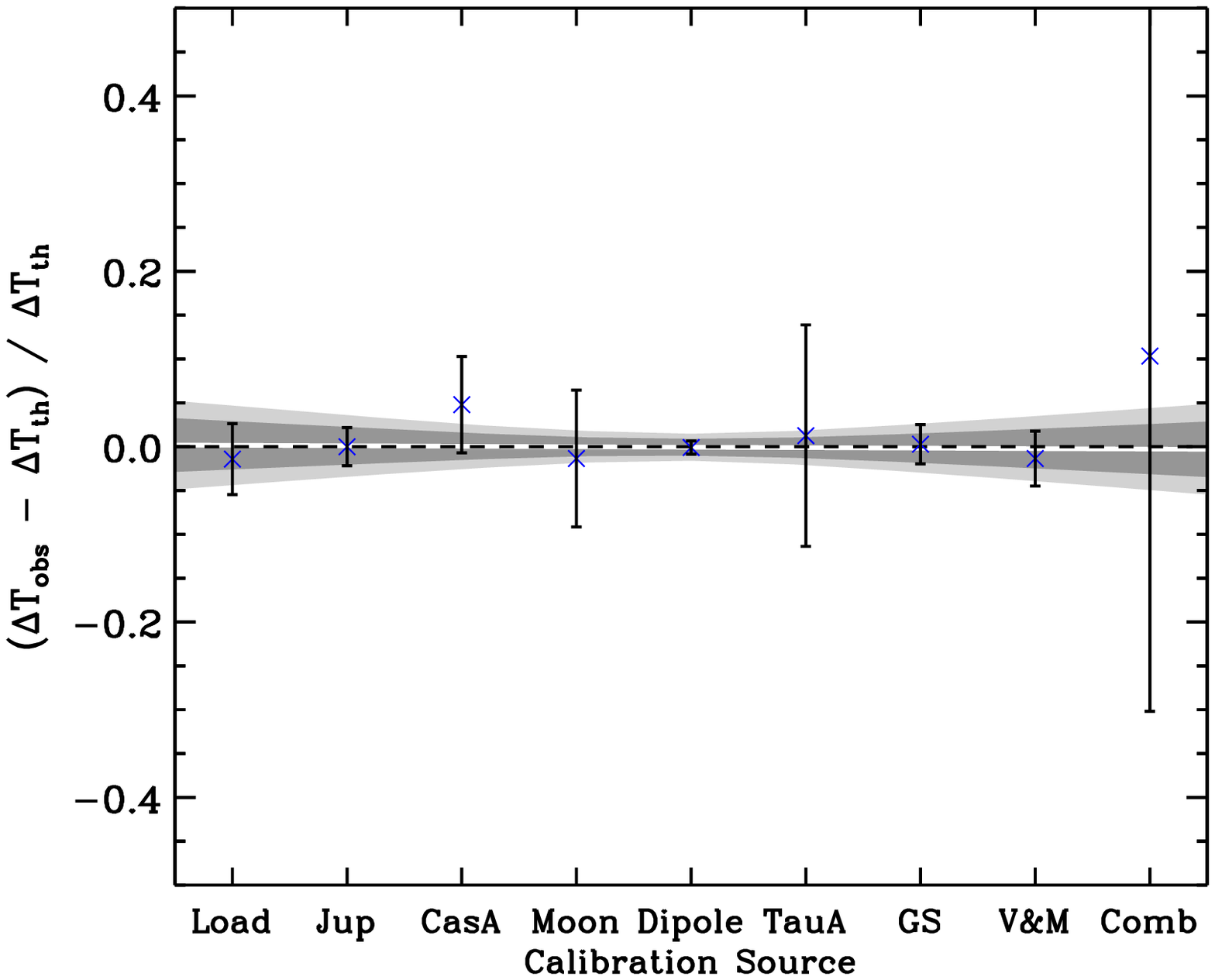,height=9.5cm,width=15cm}}
\vspace{-1.4cm}
\centerline{\psfig{figure=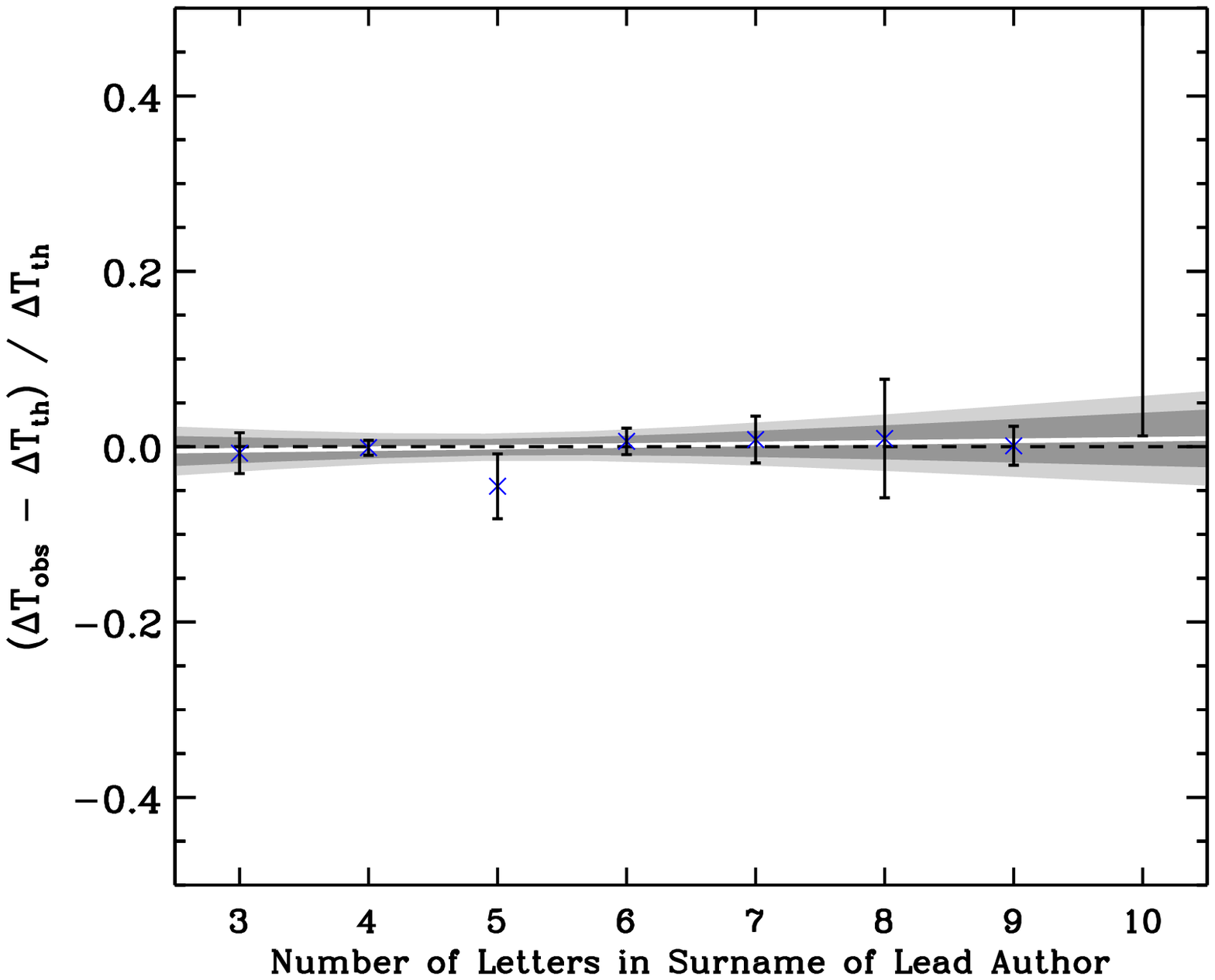,height=9.5cm,width=15cm}}
\vspace{-1.0cm}
\caption{TOP: residuals plotted against calibration
source. There seem to be no suspicious outliers.
The order of the calibration sources is arbitrary so the fitting
of a line serves only to verify that the line-fitting
and confidence-interval-determining code are working as expected.
BOTTOM: there seem to be no suspicious outliers associated  
with the number of letters in the last name of the first author.
The best-fit line here is also an exercise in checking the code.
}
\label{csresid}
\end{figure}

\section{Discussion and Summary}

We have performed linear fits on the residual data with respect to
angular size ($\ell$), galactic latitude, frequency, frequency channels, frequency bandwidth,
galactic longitude, instrument type, platform and altitude, pointing uncertainty and
area of sky observed (Griffiths \& Lineweaver 2003).
For the majority, we find little or no evidence for any trend.  
The most significant linear trend observable in the residuals is with respect
to the absolute galactic latitude $|b|$ of the observations. 
The top panel of Fig. 3 shows a linear trend that is inconsistent at more than 95\% 
confidence with a zero gradient line through the residuals. 
This trend is not eliminated by the removal of any one experiment and may be 
indicative of a source of galactic emission that has not been appropriately treated.
The best-fit line to the data suggests that CMB observations made closer to the 
galactic plane may be over-estimated by approximately 2\%.
The weighted average of points with $|b| > 40^{\circ}$ is $\sim - 1\%$, while
it is $\sim + 1\%$ for $|b| < 40^{\circ}$.  If this is due to galactic
contamination, then the normalization Q$_{10}$ may have to be reduced
$2\%$ or by $2 \sigma$ from $16.3 \; \mu$K to $16.1 \; \mu$K.
A more detailed analysis of galactic dust and synchrotron maps may reveal the source of this
trend.

At the level of $\sim 1$ to $\sim 2 \sigma$ we find  trends in the residuals with respect 
to $\ell$ (angular size).  Figure~\ref{bin_clplotcol} indicates that the
6 bins between $900\le\ell\le 1800$ prefer a lower normalization.
This could be due to underestimates of beam sizes or pointing
uncertainties or unidentified beam smearing effects at high $\ell$ for
small beams.   Although we see no significant evidence for trends associated 
with beam size or pointing uncertainty, limiting the pointing
uncertainty analysis to the 5 points with the largest uncertainties
yields a trend, suggesting that the largest pointing
uncertainties may have been underestimated.

To analyze various experiments, knowledge of the calibration
uncertainty of the measurements is necessary.  Independent
observations that calibrate off the same source will have calibration
uncertainties that are correlated at some level and therefore a
fraction of their freedom to shift upward or downward will be
shared.  
In this analysis we have marginalized over the calibration uncertainties 
associated with the observations, treating them as independent free parameters with Gaussian
distributions about their nominal values.

Over the past 10 years, successive independent and semi-independent data sets 
have extended the angular scale, calibration precision and freedom from galactic contamination
of the CMB power spectrum. The WMAP measurements (Bennett et al. 2003) are a milestone 
in this direction.
More WMAP data and the results of other CMB experiments are eagerly awaited.
Each CMB measurement contains useful cosmological information and no data set is immune
to contamination. It is therefore important to compare data sets and check for systematics.
Our results indicate that the pre-WMAP data set is consistent with
the concordant model although the model seems to need slight modifications around the scale
of the first peak and at high $\ell$ (Fig. 2).
We find no significant evidence for inter-experiment inconsistencies other than an
indication of low-level galactic contamination (Fig. 3).
We are now extending our analysis to include the WMAP data.

LMG thanks Martin Kunz for useful discussions and is
grateful to the University of Sussex where part of the work was
carried out.  LMG acknowledges support from the Royal Society and
PPARC. CHL acknowledges a research fellowship from the Australian
Research Council. 
\end{document}